\documentclass[preprint]{aastex63}

\usepackage{color}
\usepackage{multirow}
\usepackage{amsmath}

\begin{document}

\title{Various Features of the X-class White-light Flares in Super Active Region NOAA\,13664}

\author[0000-0002-8258-4892]{Ying Li}
\author[0000-0002-3657-3172]{Xiaofeng Liu}
\author[0000-0002-8401-9301]{Zhichen Jing}
\author[0000-0001-5279-3266]{Wei Chen}
\author[0000-0001-7540-9335]{Qiao Li}
\author[0000-0002-4241-9921]{Yang Su}

\affiliation{Key Laboratory of Dark Matter and Space Astronomy, Purple Mountain Observatory, CAS, Nanjing 210023, People's Republic of China}
\affiliation{School of Astronomy and Space Science, University of Science and Technology of China, Hefei 230026, People's Republic of China}

\author[0000-0003-0057-6766]{De-Chao Song}
\affiliation{Key Laboratory of Dark Matter and Space Astronomy, Purple Mountain Observatory, CAS, Nanjing 210023, People's Republic of China}

\author[0000-0002-4978-4972]{M. D. Ding}
\affiliation{School of Astronomy and Space Science, Nanjing University, Nanjing 210023, People's Republic of China}
\affiliation{Key Laboratory for Modern Astronomy and Astrophysics (Nanjing University), Ministry of Education, Nanjing 210023, People's Republic of China}

\author[0000-0003-4655-6939]{Li Feng}
\author[0000-0003-1078-3021]{Hui Li}
\affiliation{Key Laboratory of Dark Matter and Space Astronomy, Purple Mountain Observatory, CAS, Nanjing 210023, People's Republic of China}
\affiliation{School of Astronomy and Space Science, University of Science and Technology of China, Hefei 230026, People's Republic of China}

\author[0000-0001-9979-4178]{Weiqun Gan}
\affiliation{Key Laboratory of Dark Matter and Space Astronomy, Purple Mountain Observatory, CAS, Nanjing 210023, People's Republic of China}
\affiliation{University of Chinese Academy of Sciences, Nanjing 211135, People's Republic of China}

\correspondingauthor{Ying Li}
\email{yingli@pmo.ac.cn}

\begin{abstract}
Super active region NOAA 13664 produced 12 X-class flares (including the largest one, an occulted X8.7 flare, in solar cycle 25 so far) during 2024 May 8--15 and 11 of them are identified as white-light flares. Here we present various features of these X-class white-light flares observed by the White-light Solar Telescope (WST) on board the Advanced Space-based Solar Observatory and the Helioseismic and Magnetic Imager (HMI) on board the Solar Dynamics Observatory. It is found that both the white-light emissions at WST 3600 \AA\ (Balmer continuum) and HMI 6173 \AA\ (Paschen continuum) show up in different regions of the sunspot group in these flares, including outside the sunspots and within the penumbra and umbra of the sunspots. They exhibit a point-, ribbon-, loop-, or ejecta-like shape, which can come from flare ribbons (or footpoints), flare loops, and plasma ejecta depending on the perspective view. The white-light duration and relative enhancement are measured and both parameters for 3600 \AA\ emission have greater values than those for 6173 \AA\ emission. It is also found that these white-light emissions are cospatial well with the hard X-ray (HXR) sources in the on-disk flares but have some offsets with the HXR emissions in the off-limb flares. In addition, it is interesting that the 3600 and 6173 \AA\ emissions show different correlations with the peak HXR fluxes, with the former one more sensitive to the HXR emission. All these greatly help us understand the white-light flares of a large magnitude from a super active region on the Sun and also provide important insights into superflares on Sun-like stars.
\end{abstract}



\section{Introduction} 
\label{sec:intro}

Solar white-light flares (WLFs) are a special kind of flares that show an enhancement in the visible continuum. The first reported solar flare, i.e., the Carrington event in 1859 \citep{Carrington1859,Hodgson1859}, was a well-known WLF. After that, several hundreds of solar WLFs were observed in spatially resolved images and also multiwavelength spectra with ground- and space-based telescopes. It has been found that solar WLFs are similar to stellar flares in many aspects \citep[e.g.,][]{Worden1983}, the latter of which are mainly observed in light curves by Kepler and Transiting Exoplanet Survey Satellite in the optical wave band \citep[e.g.,][]{Pugh2016}. Therefore, studying solar WLFs via images in detail can provide invaluable insights into stellar flares. 

Solar WLFs are generally categorized into two types based on their observational features \citep[e.g.,][]{Fang1995,Ding2007}. Type I WLFs display enhanced Balmer continua (or a Balmer jump) and prominent Balmer lines, which are usually associated well with the hard X-ray (HXR) emissions. Type II WLFs rarely show those features, though some of them may also be accompanied by HXR emissions \citep{Prochazka2018}. The white-light (WL) emissions can be caused by a nonthermal electron-bean heating \citep[e.g.,][]{Hudson1972,Krucker2015} and its secondary effect, such as radiative backwarming \citep{Gan1994,Ding2003,Xu2004} and chromospheric condensation \citep{Gan1992,Kowalski2015}. Some other mechanisms such as a proton-beam heating \citep{Machado1978,Hiei1982,Prochazka2018}, X-ray and EUV irradiations \citep{Henoux1977,Poland1988}, and Alfv\'{e}n waves \citep{Emslie1982,Fletcher2008} can also contribute to the WL emissions in solar flares.

Super active regions (ARs) have been studied for a few solar cycles \citep[e.g.,][]{Chen2011,Le2014}, which can produce numerous flares including some largest ones in that solar cycle. The top largest flares (say, above X5.0) in a super AR are supposed to be WLFs \citep{Neidig1983}, for example, the X9.3 flare on 2017 September 6 from NOAA 12673 in solar cycle 24 \citep{Romano2018}. It should be noted that the AR that produced the Carrington WLF is a super one whose sunspot area was greater than 2400 millionths of the solar hemisphere (MSH) (the whole sunspot group being 2971 MSH) \citep{Hayakawa2023}. In the current solar cycle 25, there has appeared a super AR numbered as NOAA 13664 whose maximum area reached 2761 MSH \citep{Hayakawa2024}, i.e., comparable with the Carrington AR. NOAA 13664 produced 12 X-class and $\sim$62 M-class flares when it crossed the visible solar disk from 2024 May 1 to May 15. This region along with its 12 X-class flares were well observed by the Advanced Space-based Solar Observatory (ASO-S; \citealt{Gan2023}) and the Solar Dynamics Observatory (SDO; \citealt{Pesnell2012}). In particular, 11 of the X-class flares (excluding the first X1.0) are identified as WLFs.

In this Letter, we focus on the 12 X-class flares, especially on the 11 WLFs among them, occurred in the super AR NOAA 13664 during 2024 May 8--15. Thanks to the routine full-disk imaging observations from the White-light Solar Telescope (WST) on the Ly$\alpha$ Solar Telescope (LST; \citealt{LiH2019,Feng2019,Chen2024}) on board ASO-S and the Helioseismic and Magnetic Imager (HMI; \citealt{Scherrer2012}) on board SDO, we can study the emission features of these X-class WLFs at 3600 and 6173\,\AA\ (i.e., below and above the Balmer limit) simultaneously. Properties of the WL emissions including their morphology, source, duration, and enhancement are presented in this study, together with their physical origins. All these can significantly improve our understanding on the solar WLFs and further provide important insights into superflares on the Sun-like stars.


\section{Observational Data}
\label{sec:data}

The WL continuum data used in this work are obtained by SDO/HMI and ASO-S/WST. WST provides the full-disk imaging data at 3600 \AA\ (in the Balmer continuum) with a cadence of two minutes routinely. The images have a pixel size of $\sim$0.5\arcsec\ whereas the spatial resolution is $\sim$4\arcsec\ due to a poor point spread function. HMI obtains the magnetogram data via the \ion{Fe}{1} line at 6173 \AA\ as well as pseudo-continuum (Paschen continuum) images near the line which have a pixel size of 0.5\arcsec\ and a cadence of 45 s. We also use the HXR data observed by the Hard X-ray Imager (HXI; \citealt{Zhang2019,Su2022}) on ASO-S and some EUV/UV images from the Atmospheric Imaging Assembly (AIA; \citealt{Lemen2012}) on SDO. HXI observes the full Sun in an energy range of $\sim$10--300 keV. Its energy and angular resolutions are $\sim$16.5\% at 32 keV and 3.1\arcsec, respectively. AIA provides the EUV/UV images at 131, 94, 335, 193, 211, 171, 304, 1600, and 1700 \AA\ with a cadence of 12 or 24 s. These images have a pixel size of 0.6\arcsec\ and the spatial resolution is 1.2\arcsec. The soft X-ray (SXR) flux at 1--8 \AA\ observed by the X-Ray Sensor (XRS; \citealt{Hanser1996}) on the Geostationary Operational Environmental Satellite (GOES) is presented in this work as well, which is used to divide the flares into A, B, C, M, and X classes.


\section{Observations and Results}
\label{sec:result}

\subsection{Overview of the 12 X-class Flare Events in Super AR NOAA 13664}

\begin{figure}[htb]
	\centering
	\includegraphics[width=\textwidth]{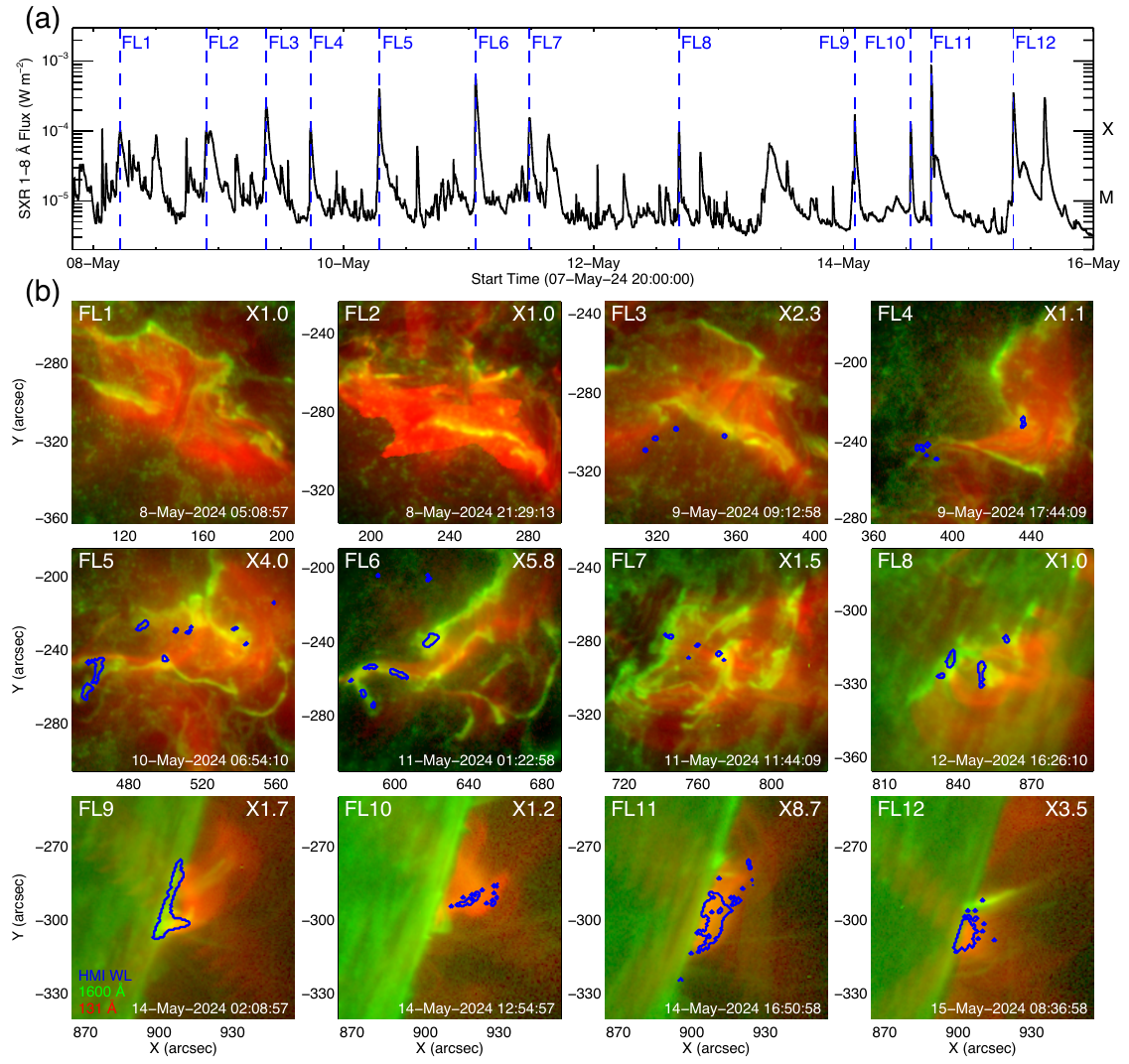}
	\caption{Overview of the 12 X-class flare events in super AR NOAA 13664. (a) GOES SXR flux during 2024 May 7--16. The blue vertical dashed lines in the panel denote the peak times of the 12 X-class flares. (b) AIA 1600 \AA\ (green) and 131 \AA\ (red)  images around the peak times of the 12 flares. The blue contours in the images mark the HMI 6173 \AA\ brightenings, which are the same as the red contours in Figure \ref{fig2}(a).}
	\label{fig1}
\end{figure}

The 12 X-class flares, labelled as FL1 to FL12 (see Table \ref{tab:list}), occurred in the super AR NOAA 13664 during 2024 May 8--15. NOAA 13664 appeared at the solar east limb on May 1, reached a sunspot area of $\sim$2400 MSH on May 11 with a $\beta\gamma\delta$/Fkc sunspot group, and rotated over the west limb on May 15. This region produced numerous flares (including the largest X8.7 in the solar cycle 25 so far) as seen from the GOES SXR 1--8\,\AA\ light curve in Figure \ref{fig1}(a). Figure \ref{fig1}(b) shows the AIA 1600 and 131\,\AA\ images around the peak times of these X-class flares. One can see that most of the events (i.e., on-disk FL1--FL8) show clear ribbons and hot flaring loops. When the region rotated over the west limb, only flare loops were visible above the limb (in the off-limb FL10--FL12).

Among the 12 X-class flares, 11 are identified as WLFs (see the remarks in Table \ref{tab:list}), and only the first X1.0 flare (FL1) is not a detectable WLF (NWLF). Here we define a WLF using intensity thresholds (a ratio of the intensity at a single flaring pixel and the average intensity over a quiet-Sun region near the flaring region) of 5\% for HMI 6173 \AA\ emissions \citep{Song2018} and 8\% for WST 3600 \AA\ emissions \citep{Jing2024} for the on-disk flares, i.e. FL1--FL8. These threshold values are about three times the intensity fluctuations or standard deviations of the intensity over the nearby quiet-Sun region during the whole flare time, which are within 1.4\% and 2.9\% for HMI and WST, respectively. Note that the second X1.0 flare (FL2) shows some reliable brightening at WST 3600 \AA\ (marked by the red arrow in Figure \ref{fig3}(b)) but its intensity ratio is actually below the above thresholds. For the on- and off-limb events of FL9--FL12, since their WL brightenings are relatively weak, we just identify them as WLFs via checking the base-difference images (see Figure \ref{fig3}) by eye. From Figure \ref{fig2} it is seen that the 3600 \AA\ brightening can be visible in the original WST images for some large or  on-limb flares, say, FL5 (X4.0), FL6 (X5.8), and FL9 (X1.7), while in the other WLFs the 3600 and 6173\,\AA\ brightenings can only be seen from the base- or running-difference images as shown in Figure \ref{fig3}. 

\subsection{Properties of the WL Emissions at WST 3600\,\AA\ and HMI 6173\,\AA}

\subsubsection{WL Morphology and Source}

\begin{figure}[htb]
	\centering
	\includegraphics[width=\textwidth]{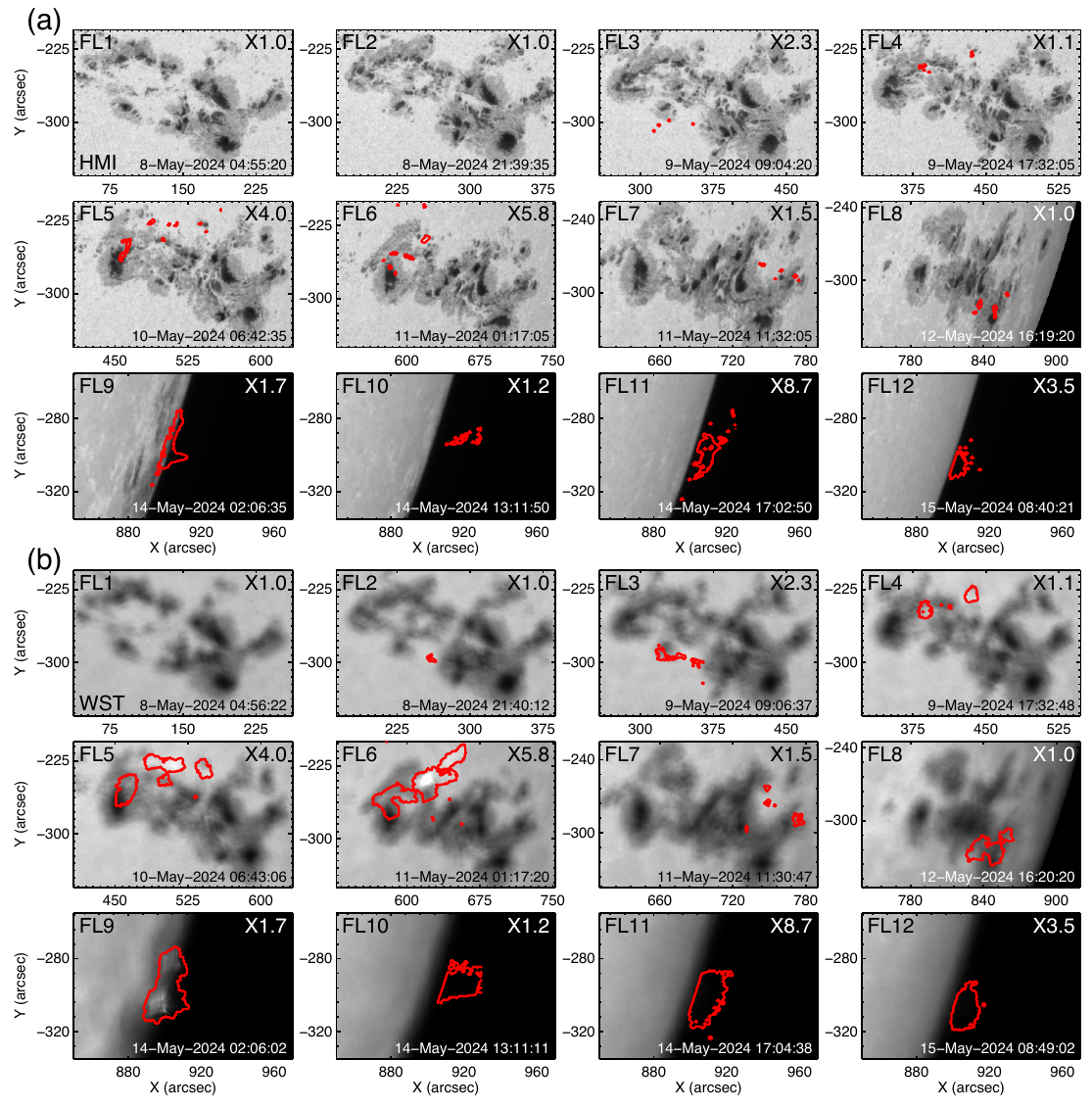}
	\caption{HMI (a) and WST (b) images at the WL brightening time for the 12 X-class flares. The red contours with fixed levels for on-disk (except FL3 and FL7 having a weak WL emission) and off-limb flares respectively in the images indicate the WL brightenings that correspond to those as seen in the running- or base-difference images in Figure \ref{fig3}. Note that FL1 is not a detectable WLF without showing the WL contours and that FL2 without the HMI contour is due to a data missing during its WL brightening period.}
	\label{fig2}
\end{figure}

\begin{figure}[htb]
	\centering
	\includegraphics[width=\textwidth]{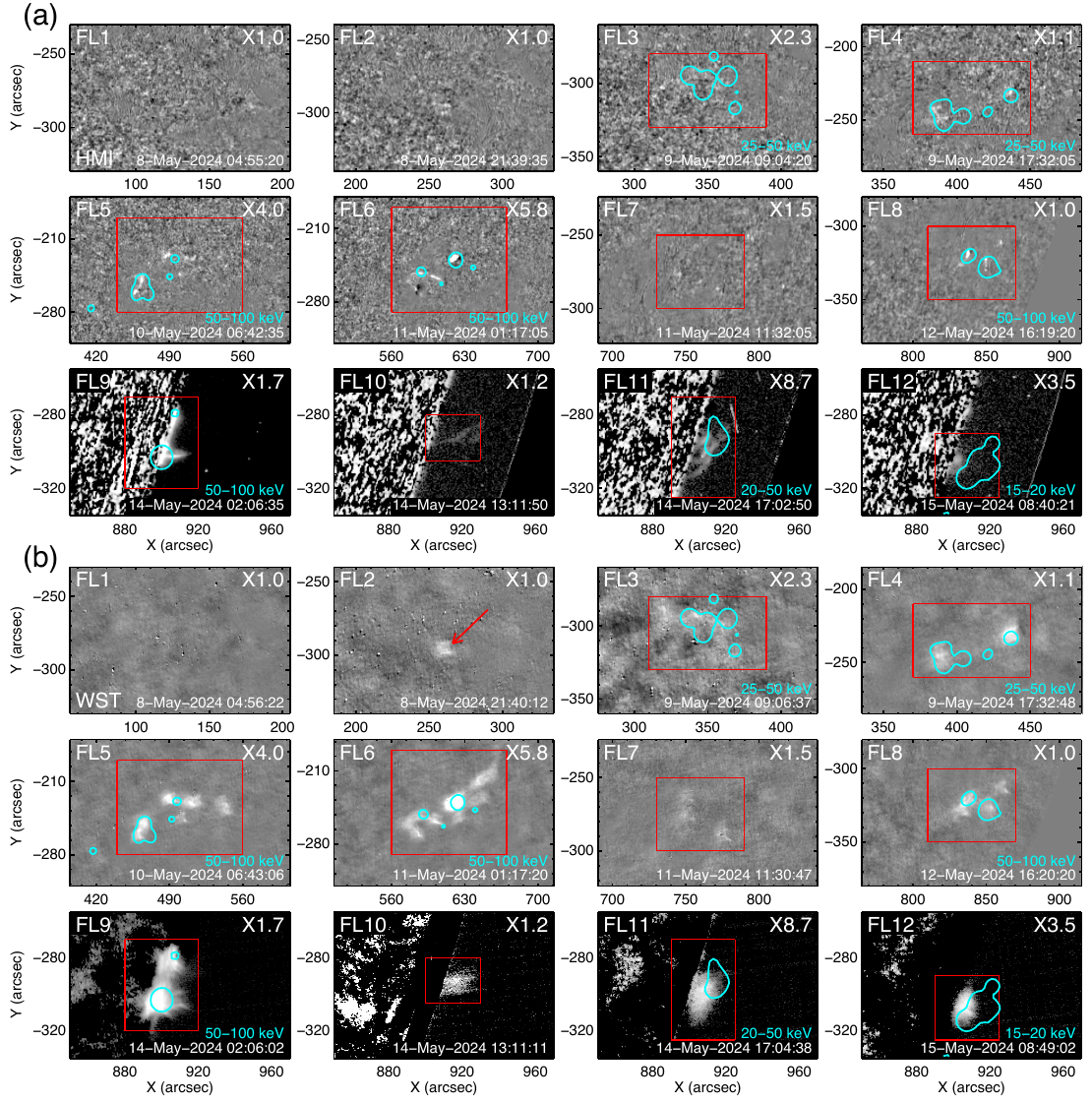}
	\caption{HMI (a) and WST (b) running- (for on-disk flares) or base-difference (for off-limb flares) images at the WL brightening time for the 12 X-class flares. The cyan contours (with 20\% level at the maximum) in the images mark the HXI HXR sources for the WLFs that have available HXI data at the corresponding time. The red boxes denote the regions integrated over for the HMI and WST emission curves in Figure \ref{fig4}. The red arrow in the WST image of FL2 indicates the visible 3600 \AA\ brightening though its enhancement is below our threshold.}
	\label{fig3}
\end{figure}

From Figure \ref{fig2} (the red contours marking the WL brightenings) one can see that the bright 6173 and 3600\,\AA\ emissions show up in different regions of the sunspot group. In some WLFs, these two wave-band brightenings mainly appear outside the sunspots, say, in FL2 and FL3. By contrast, the brightenings can be located in the penumbra or even umbra of the sunspots in FL4--FL9. In the off-limb WLFs of FL10--F12 with their foopotints occulted by the solar limb, the WL emissions are mainly seen above the limb or in the corona.

From Figure \ref{fig3} (and also the red contours in Figure \ref{fig2}) we can see that the bright emissions at 6173 and especially 3600\,\AA\ display various shapes in the morphology (also listed in Table \ref{tab:wl-info}). In FL2--FL4, FL7, and FL8 (all $<$X3.0), the WL brightenings mainly show a point-like shape, say, a single point (FL2), double points (FL4), and even three points (FL7 and FL8). By contrast, the bright emissions in FL5 and FL6 ($\ge$X4.0) show a ribbon-like shape. Note that the brightening area and also the enhancement at 6173\,\AA\ are smaller than those at 3600\,\AA\ (see the following Section). These on-disk WL emissions in FL2--FL8 are mainly located at the flare footpoints or ribbons combining with the AIA 1600\,\AA\ images in Figure \ref{fig1}(b). Note that the times of the WL and 1600 \AA\ emissions are not exactly the same. In the remaining four on- or off-limb flares (FL9--FL12), the WL emissions exhibit a loop- or ejecta-like shape, corresponding to the flare loop structures in the corona as seen in AIA 131\,\AA\ images in Figure \ref{fig1}(b). Note that in the on-limb FL9, the WL brightening is also from chromospheric footpoints.

\subsubsection{WL Enhancement and Duration}

The relative enhancement of WL emission is defined as $(I-I_{ref})/I_{ref}$, where $I$ and $I_{ref}$ are the WL intensity during the flare and the reference intensity, respectively. For WST, we adopt the average intensity over 10 minutes during the preflare phase as $I_{ref}$, while the average intensity over the 15 to 9 minutes before the WL peak time is used as $I_{ref}$ for HMI in order to avoid the sunspot motion effect as much as possible. We provide the enhancements of both HMI 6173\,\AA\ and WST 3600\,\AA\ emissions averaged over the maximum WL area for the on-disk WLFs, as listed in Table \ref{tab:wl-info}. Here the uncertainties or errors are derived by varying the intensity thresholds of identifying a WLF, say, 5$\pm$1.02\% and 8$\pm$1.26\%, where 1.02\% and 1.26\% are the averaged standard deviations of the intensity fluctuations over a quiet-Sun region for HMI and WST, respectively. It is seen that the enhancement of 3600\,\AA\ emission is greater than that of 6173\,\AA\ in all FL3--FL8, i.e., type I WLFs basically. The former has a range of 13--27\% while the latter is 9--15\%. These enhancements are similar to those reported in \cite{Song2018} and \cite{Jing2024} though most of the WLFs in both studies are M-class ones. For the on- and off-limb WLFs, they could be referred to as coronal type (listed in Table \ref{tab:wl-info}) due to the WL emissions appearing in the corona.

\begin{figure}[htb]
	\centering
	\includegraphics[width=\textwidth]{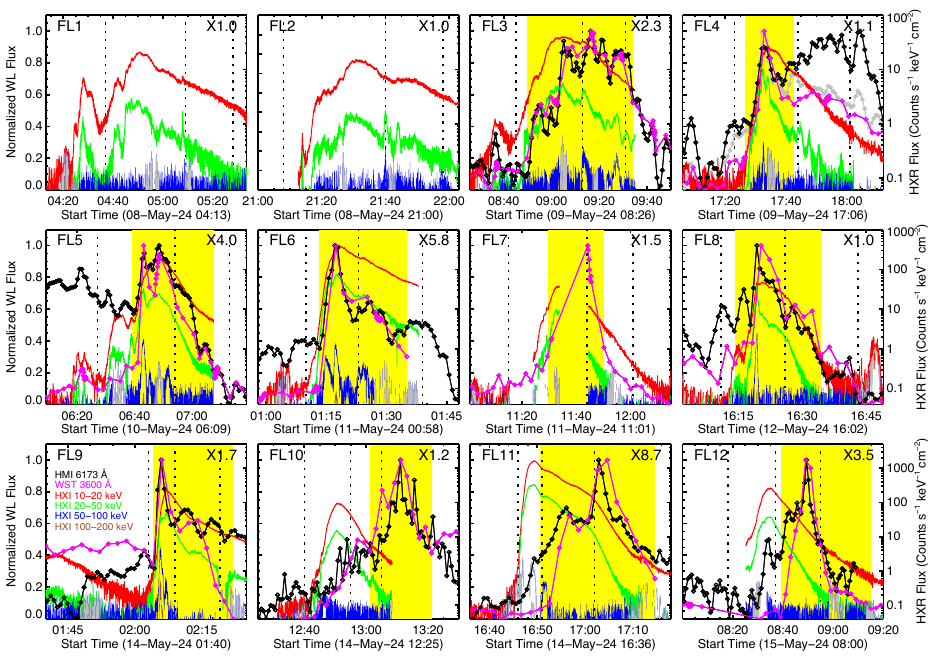}
	\caption{HXR and WL emission curves from HXI, HMI, and WST for the 12 X-class flares. The HMI and WST emissions are integrated over the same flaring region marked by the red box in Figure \ref{fig3}. For FL4, we also overplot the HMI emission curve (in gray) integrated over the compact WL sources only for reference. In each panel, the three vertical dotted lines indicate the flare onset, peak, and end times. The yellow shaded area denotes the duration of the WL emission at WST 3600 \AA. Note that the time periods affected by the radiation belt are indicated in gray in the HXR 50--100 keV curve and that the poor HXR data during the times of the South Atlantic Anomaly (SAA) and HXI night are not displayed.}
	\label{fig4}
\end{figure}

The WL duration is measured from the WST plus HMI light curve. Figure \ref{fig4} plots the normalized WL curves at 6173 and 3600\,\AA\ integrated over the flaring region marked by the red box in Figure \ref{fig3} for FL3--FL12. Note that FL7 exhibits a very weak brightening at 6173\,\AA, which is not obvious in the integrated light curve, thus the HMI curve is not shown here. It is seen that the two wave-band emission curves display a very similar trend in majority of the WLFs. Based on these WL curves, together with the WST images in which the brightening is easier to be seen, we determine the WL duration just by eye. The results are shown with a yellow shaded area in Figure \ref{fig4} and also listed in Table \ref{tab:wl-info}. One can see that except FL2, all the identified WLFs have a duration of longer than 15 minutes with a range of $\sim$16--44 minutes. Note that the durations determined by eye here are likely to be overestimated for the HMI 6173\,\AA\ emission which usually exhibits a weaker enhancement and has a smaller brightening area. Our results are indeed greater than the durations (only about five minutes) reported in \cite{Song2018} for 6173\,\AA\ emissions. It should also be mentioned that our durations are fairly similar to those measured by \cite{Jing2024} for 3600\,\AA\ emissions in M-class and especially X-class WLFs.

\subsubsection{Relationship of the WL brightening with the HXR Emission}

Firstly, we check the spatial relationship between the WL and HXR emissions. The HXR images are obtained via the HXI CLEAN algorithm \citep{Su2019} and their contours are overplotted on the running- or base-difference images of HMI and WST in Figure \ref{fig3}. Note that the HXR images of FL7 and FL10 are not provided mainly due to the corresponding HXR data being unavailable at that time. From Figure \ref{fig3} one can see that in the on-disk/limb WLFs (FL3--FL6, FL8, and FL9), the WL brightenings are cospatial with the HXR footpoint sources well. For the two off-limb WLFs (FL11 and FL12), the WL emissions are mostly  under the HXR looptop sources with some offsets. These will be discussed in the following Section.

Then we check the temporal relationship between the WL and HXR emissions, which shows varies cases. Figure \ref{fig4} plots the background-subtracted HXR light curves at 10--20, 20--50, 50--100, and 100--200\,keV from HXI. One could see that in FL3, the 6173 and 3600\,\AA\ emissions seem to peak later than the HXR emissions at 10--20 and 20--50\,keV and even later than the SXR 1--8\,\AA\ emission. Here it should be noted that the HMI and WST emission curves for this flare show an oscillation pattern and the images look somewhat noise-like (Figure \ref{fig3}). Moreover, the current cadences of HMI and WST might not be high enough to determine the true peaks. For FL4, the HMI and WST emissions seem to reach the maximum around the same time as the HXR emission at 20--50\,keV and before the SXR emission. Note that here we overplot the HMI emission curve integrated over the WL sources only for reference (see the gray curve). In FL5, FL6, FL8, and FL9, the WL emissions peak around the same time as the HXR 50--100\,keV emission (and also HXR 100--200\,keV) in the rise phase of the flare. For the off-limb FL10--FL12, their WL emissions reach the maximum during the decay phase and much later than the HXR emissions mainly due to the footpoints being occulted.

Finally, it is interesting to find that the WL enhancements have some good correlations with the peak SXR as well as HXR fluxes (listed in Table \ref{tab:wl-info}) and particularly that the 6173 and 3600 \AA\ emissions show different dependences on the HXR emissions. In FL3--FL8, their 3600 \AA\ enhancements have good correlations with their peak HXR fluxes at 20--50 and 50--100 keV, with correlation coefficients (cc) of 0.73 and 0.91, respectively. While the 6173 \AA\ enhancements only show a good correlation (cc$=$0.91) with the peak HXR flux at 50--100 keV. Moreover, the enhancement ratios of 3600 and 6173 \AA\ emissions have some relationships with the peak HXR fluxes at 20--50 keV (cc$=$0.98) and 50--100 keV (cc$=$0.59) and also peak SXR flux (cc$=$0.76) in these WLFs.


\section{Summary and Discussions}
\label{sec:summary}

In this study, we present various features of 11 X-class WLFs observed by ASO-S and SDO during 2024 May 8--15, which occurred in the super AR NOAA 13664 in the current solar cycle 25. Thanks to the routine full-disk imaging observations of ASO-S/WST at 3600 \AA\ and SDO/HMI at 6173 \AA, we can investigate the Balmer and Paschen continuum emissions simultaneously and compare with each other in these large magnitude WLFs. Our results are summarized as follows.

\begin{enumerate}
\item 11 out of the 12 X-class flares (except the first X1.0) are identified as WLFs. Their bright 6173 and 3600 \AA\ emissions show up in different regions of the sunspot group, including outside the sunspots and within the penumbra and umbra of the sunspots.
\item The 6173 and 3600 \AA\ brightenings exhibit various shapes such as point-, ribbon-, loop-, and ejecta-like and can come from flare footpoints/ribbons mainly rooted in the chromosphere (some might be deep into the photosphere) and also from flare loops in the corona. 
\item The relative enhancement of 3600 \AA\ emissions is greater than that of 6173 \AA\ emissions for the on-disk WLFs, with the former in a range of 13--27\% and the latter of 9--15\%. The duration of 3600 \AA\ emissions ranges from 16 to 44 minutes, which is supposed to be longer than that of 6173 \AA\ emissions.
\item The 3600 and 6173 \AA\ brightenings are cospatial well with the HXR sources in the on-disk WLFs, which, however, have some offsets with the HXR emissions in the off-limb WLFs. These WL emissions can peak around the same time or (much) later than the HXR emission. It is also interesting that the 3600 and 6173 \AA\ emissions show different dependences on the peak HXR fluxes, with the former one more sensitive to the HXR emission.
 \end{enumerate}
 
According to the spatial relationship between the WL and HXR emissions, we speculate that the 3600 and also 6173 \AA\ emissions at flare footpoints are closely related to an electron-beam heating \citep[e.g.,][]{Song2023} that mainly happens in the rise phase of the flare. This can explain a good match of the peak times of WL and HXR emissions in some of the WLFs. The secondary effect of the beam heating such as radiative backwarming can also contribute to the 3600 and especially 6173 \AA\ emissions \citep[e.g.,][]{Hao2017}, which usually causes a delay between the WL and HXR emissions in some of the other WLFs. As regards the 3600 and 6173 \AA\ emissions on flare loops during the decay phase, which show a spatial offset with the HXR sources, they are supposed to be related to thermal plasma cooling \citep[e.g.,][]{ying2024} or due to Thomson scattering \citep[e.g.,][]{Hiei1992,Saint2014,Heinzel2017}. Note that here we could not exclude that cool plasma ejecta generates the WL emissions above the limb \citep[e.g.,][]{Oliveros2014,Saint2014,Fremstad2023}. Overall, the relationships between the WL and HXR emissions show various cases, thus revealing various mechanisms responsible for the WL emissions in these X-class WLFs.

The 11 X-class WLFs together with the first non-WLF (or NWLF) exhibit some interesting developing trends as the super AR evolves over time. (1) The first X1.0 flare is not a detectable WLF, while the second X1.0 flare begins to show a short and visible 3600 \AA\ brightening (still below our threshold). Then all the following X-class flares exhibit 3600 and 6173 \AA\ emissions above the thresholds. (2) The WL emissions at 6173 and 3600 \AA\ appear outside the sunspots (or in a quiet-Sun region) in the first few WLFs, which begin to show up within the penumbra and even umbra in the following WLFs. (3) The HXR emissions can only be visible mainly at 20--50 keV in the first few flares, which become notable at 50--100 keV and even 100--200 keV in the following on-disk or on-limb flares. (4) As the AR rotated over the west limb, the WL emissions can be seen from the chromospheric footpoints to flare loops in the corona. The former three should be related to the time evolution of the magnetic and sunspot complexity of this super AR, which is worthwhile to be studied in the future.

 The routine full-disk imaging observations of WST and HMI give us good opportunities to systematically investigate the Balmer and Paschen continuum emissions at 3600 and 6173 \AA\ respectively in WLFs. In particular, the 3600 and 6173 \AA\ emissions show various features in the 11 X-class WLFs from this single super AR. These provide important constrains for the radiative hydrodynamic modeling of solar and stellar flares, especially combining with spectroscopic observations from the Chinese H$\alpha$ Solar Explorer (CHASE; \citealt{LiC2022}) and the Interface Region Imaging Spectrograph (IRIS; \citealt{DePontieu2014}), which will help us well understand the energy transportation and deposition of WLFs \citep{Heinzel2014,Kleint2016}.

\acknowledgments
We appreciate the referee's valuable suggestions and comments that helped to improve our manuscript. The ASO-S mission is supported by the Strategic Priority Research Program on Space Science, Chinese Academy of Sciences. SDO is a mission of NASA's Living With a Star Program. The authors are supported by the Strategic Priority Research Program of the Chinese Academy of Sciences under grant XDB0560000, the National Key R\&D Program of China under grant 2022YFF0503004, and NSFC under grants 12273115, 12233012, and 11921003.

\bibliographystyle{apj}

\begin{table}[htb]
\caption{Information of the 12 X-class Flares from Super AR NOAA 13664}
\label{tab:list}
\centering
\begin{tabular}{lccccccc}
\hline
\hline
Flare~~ & ~~Observation~~ & ~~GOES~~ & ~~GOES~~ & ~~GOES~~ & ~~GOES~~ & AR & \\
No. & Date & Start & Peak & End & Class & ~~Location~~ & ~~Remarks \\
 & (yyyy-mm-dd) & (UT) & (UT) & (UT) & & & \\ 
\tableline
FL1 & 2024-05-08 & 04:37 & 05:09 & 05:28 & X1.0 & S22W11 & on-disk, NWLF \\
FL2 & 2024-05-08 & 21:08 & 21:40 & 21:58 & X1.0 & S20W17 & on-disk, ``WLF"$^{a}$ \\ 
FL3 & 2024-05-09 & 08:45 & 09:13 & 09:31 & X2.3 & S20W24 & on-disk, WLF \\ 
FL4 & 2024-05-09 & 17:23 & 17:44 & 18:01 & X1.1 & S17W28 & on-disk, WLF \\ 
FL5 & 2024-05-10 & 06:27 & 06:54 & 07:13 & X4.0 & S17W34 & on-disk, WLF \\ 
FL6 & 2024-05-11 & 01:10 & 01:23 & 01:39 & X5.8 & S17W44 & on-disk, WLF \\ 
FL7 & 2024-05-11 & 11:15 & 11:44 & 12:01 & X1.5 & S19W60 & on-disk, WLF \\ 
FL8 & 2024-05-12 & 16:11 & 16:26 & 16:43 & X1.0 & S20W75 & on-disk, WLF \\ 
FL9 & 2024-05-14 & 02:03 & 02:09 & 02:19 & X1.7 & S17W89 & on-limb, WLF \\
FL10 & 2024-05-14 & 12:40 & 12:55 & 13:05 & X1.2 & S17W89 & off-limb, WLF \\ 
FL11 & 2024-05-14 & 16:46 & 16:51 & 17:02 & X8.7 & S18W89 & off-limb, WLF \\ 
FL12 & 2024-05-15 & 08:18 & 08:37 & 08:55 & X3.5 & S18W89 & off-limb, WLF \\ 
\hline
\hline
\end{tabular}
$^{a}$``WLF" means that this flare has some detectable brightening at 3600 \AA\ but its relative enhancement \\is below our threshold of 8\%. 
\end{table}

 \begin{table}[htbp]
\caption{Properties of the WL Emissions for the 11 WLFs along with Their Peak HXR Fluxes}
\label{tab:wl-info}
\centering
\tiny
\begin{tabular}{lcccccccccccc}
\hline
\hline
Flare & GOES & WL Morphology & WL Source & WL Duration & ~ & \multicolumn{2}{l}{WL Enhancement$^{a}$ (\%)} & ~ & WL Ratio & Type & ~~$F_{p,20-50}$$^{c}$ & $F_{p,50-100}$$^{d}$ \\ \cline{7-8}
No. & Class &  &  & (min) & & HMI 6173 \AA\ & WST 3600 \AA\ & ~ & (3600/6173) & & \multicolumn{2}{c}{(counts s$^{-1}$ keV$^{-1}$ cm$^{-2}$)} \\ \cline{11-12}
\tableline
FL2 & X1.0 & point-like & footpoints & $\le$2 & ~ & -$^{b}$ & $<$8 & ~ & - & - & 1.77 & $\sim$0.35 \\
FL3 & X2.3 & point-like & footpoints & $\sim$44 & ~ & 10.9$\pm$2.17 & 16.9$\pm$2.59 & ~ & 1.56$\pm$0.57 & type I & 5.63 & $\sim$0.61 \\
FL4 & X1.1 & point-like & footpoints & $\sim$16 & ~ & 9.46$\pm$1.97 & 14.1$\pm$1.68 & ~ & 1.49$\pm$0.51 & type I & 10.8 & $\sim$0.47 \\
FL5 & X4.0 & ribbon-like & footpoints & $\sim$28 & ~ & 11.3$\pm$2.32 & 18.9$\pm$1.61 & ~ & 1.68$\pm$0.51 & type I & 33.0 & 1.75 \\
FL6 & X5.8 & ribbon-like & footpoints & $\sim$22 & ~ & 10.9$\pm$1.86 & 22.7$\pm$3.49 & ~ & 2.08$\pm$0.69 & type I & 74.0 & 1.37 \\
FL7 & X1.5 & point-like & footpoints & $\sim$21 & ~ & 9.27$\pm$1.43 & 13.5$\pm$2.59 & ~ & 1.45$\pm$0.52 & type I & $\sim$10.0 & $\sim$0.44 \\
FL8 & X1.0 & point-like & footpoints & $\sim$20 & ~ & 15.0$\pm$4.19 & 26.5$\pm$3.14 & ~ & 1.77$\pm$0.77 & type I & 38.8 & 2.49 \\
FL9 & X1.7 & point- \& loop-like & footpoints \& loops & $\sim$18 & ~ & - & - & ~ & - & coronal & - & - \\
FL10 & X1.2 & loop-like & loops & $\sim$20 & ~ & - & - & ~ & - & coronal & - & - \\
FL11 & X8.7 & loop-like & loops & $\sim$24 & ~ & - & - & ~ & - & coronal & - & - \\
FL12 & X3.5 & loop-like & loops & $\sim$36 & ~ & - & - & ~ & - & coronal & - & - \\
\hline
\hline
\end{tabular}
~~$^{a}$The WL enhancement refers to the one averaged over the maximum WL area. \\
$^{b}$The HMI data during the WL brightening period ($\sim$21:39--21:44 UT) were missing. \\
$^{c}$$F_{p,20-50}$ represents the peak HXR flux at 20--50 keV. \\
$^{d}$$F_{p,50-100}$ represents the peak HXR flux at 50--100 keV.
\end{table}

\end{document}